\documentclass[prb,twocolumn,showpacs,superscriptaddress]{revtex4}

\usepackage[dvips]{graphicx}
\usepackage{dcolumn}
\usepackage{bm}
\usepackage{color}
\usepackage{amssymb,amsmath}

\begin{document}


\title{Spin relaxation in Cs$_2$CuCl$_{4-x}$Br$_x$}

\author{R.~HassanAbadi}
\affiliation{Experimentalphysik V, Center for Electronic Correlations and Magnetism, Institute for Physics, Augsburg University, D-86135 Augsburg, Germany}

\author{R.~M.~Eremina}
\affiliation{Zavoisky Physical-Technical Institute, Federal Research Center "Kazan Scientific Center of RAS", 420029 Kazan, Russia}
\affiliation{Institute for Physics, Kazan (Volga region) Federal University, 420008 Kazan, Russia}

\author{M.~Hemmida}
\affiliation{Experimentalphysik V, Center for Electronic Correlations and Magnetism, Institute for Physics, Augsburg University, D-86135 Augsburg, Germany}

\author{ A.~Dittl}
\affiliation{Experimentalphysik V, Center for Electronic Correlations and Magnetism, Institute for Physics, Augsburg University, D-86135 Augsburg, Germany}

\author{ B.~Wolf}
\affiliation{Physikalisches Institut, Goethe-Universit\"{a}t Frankfurt, Max-von-Laue-Strasse 1, 60438 Frankfurt am Main,
Germany}

\author{W.~Assmus}
\affiliation{Physikalisches Institut, Goethe-Universit\"{a}t Frankfurt, Max-von-Laue-Strasse 1, 60438 Frankfurt am Main,
Germany}

\author{A.~Loidl}
\affiliation{Experimentalphysik V, Center for Electronic Correlations and Magnetism, Institute for Physics, Augsburg University, D-86135 Augsburg, Germany}

\author{H.-A.~Krug von Nidda}
\affiliation{Experimentalphysik V, Center for Electronic Correlations and Magnetism, Institute for Physics, Augsburg University, D-86135 Augsburg, Germany}

\date{\today}

\begin{abstract}

The quantum-spin $S=1/2$ chain system Cs$_2$CuCl$_4$ is of high interest due to competing antiferromagnetic intra-chain $J$ and inter-chain exchange $J^{\prime}$ interactions and represents a paramount example for Bose-Einstein condensation of magnons [R. Coldea \textit{et al.}, Phys. Rev. Lett. \textbf{88}, 137202 (2002)]. Substitution of chlorine by bromine allows tuning the competing exchange interactions and corresponding magnetic frustration. Here we report on electron spin resonance (ESR) in single crystals of Cs$_2$CuCl$_{4-x}$Br$_x$ with the aim to analyze the evolution of anisotropic exchange contributions. The main source of the ESR linewidth is attributed to the uniform Dzyaloshinskii-Moriya interaction. The vector components of the Dzyaloshinskii-Moriya interaction are determined from the angular dependence of the ESR spectra using a high-temperature approximation. The obtained results support the site selectivity of the Br substitution suggested from the evolution of lattice parameters and magnetic susceptibility dependent on the Br concentration.

\end{abstract}

\maketitle

\section{Introduction}

Low dimensional magnets exhibit a large variety of fascinating ground states, especially in case of competing interactions and geometric frustration.\cite{Vasiliev2018,Balents2010} All these factors are present in the compounds Cs$_2$CuCl$_4$ and Cs$_2$CuBr$_4$,\cite{Mellor1939, Helmholz1952,Morosin1960} where antiferromagnetic spin $S=1/2$ chains of Cu$^{2+}$ ions form triangular antiferromagnetic planes with neighboring chains. Thereby, the inter-chain interaction $J^{\prime}$ is of the same order of magnitude but always smaller than the intra-chain interaction $J$.\cite{Coldea2001,Starykh2007} In Cs$_2$CuCl$_4$ the ratio $J^{\prime}/J \approx 1/3$ results in formation of a spin-liquid state below about 4\,K. Finally, weak inter-layer couplings give rise to long-range spin-spiral order below the N\'{e}el temperature $T_{\rm N}=0.62$\,K.\cite{Coldea2001} For Cs$_2$CuBr$_4$ the ratio $J^{\prime}/J$ is larger than in the Cl system indicating a stronger frustration: the field dependence of the magnetization reveals two plateaux.\cite{Ono2004} The antiferromagnetic ground state below $T_{\rm N} = 4$\,K is interpreted in terms of triplet crystallization.\cite{Ono2003} Notably, the discovery of Bose-Einstein condensation of magnons in Cs$_2$CuCl$_4$ in magnetic fields above 8\,T triggered a permanent interest in these compounds to date.\cite{Coldea2002,Radu2005}

Spinon and magnon excitations in Cs$_2$CuCl$_4$ have been studied by magnetic resonance techniques.\cite{Vachon2011,Smirnov2012,Smirnov2015} Ultrasound experiments concentrate on the spin-liquid phase and quantum critical behavior. \cite{Streib2015,Cong2016} Very recently a change of the magnetic properties has been reported after application and release of hydrostatic pressure,\cite{Kim2017} indicating a deformation of the local environment of the copper ions. In Cs$_2$CuBr$_4$ the magnetic phase diagram of incommensurate and commensurate structures has been investigated experimentally and theoretically.\cite{Fujii2007,Alicea2009} Recent ultrasound experiments up to magnetic fields of 50 T also suggested the existence of a spin-liquid phase,\cite{Wolf2013} which was supported by the finding of a large zero-field gap even above $T_{\rm N}$ by high-field electron spin resonance.\cite{Zvyagin2015}

Chemical substitution of chlorine by bromine allows systematically tuning the exchange coupling between the Cu chains and the strength of frustration.\cite{Cong2011} Crystallographic and magnetic studies of single crystals of the series Cs$_2$CuCl$_{4-x}$Br$_x$ with $0 \leq x \leq 4$ revealed three different regimes dependent on the bromine concentration $x$ due to site-selective substitution, as there are three inequivalent Cu--Cl bonds in the distorted CuCl$_4$ tetrahedra. Following the notation of Ref.~\onlinecite{Cong2011} one finds the longest Cu--Cl1 bond tilted from the $a$ direction, the intermediate Cu--Cl2 bonds tilted from the $c$ direction, and two short Cu--Cl3 bonds involved in the formation of the Cu chains along the $b$ axis. Thus, the larger Br$^-$ anions first substitute the Cl1$^-$ anions with the largest space for $0 \leq x \leq 1$, then the Cl2$^-$ anions for $1 \leq x \leq 2$, and finally the Cl3$^-$ anions within the chains for $2 \leq x \leq 4$. Hence, regarding the $x$-dependence of the temperature $T_{\rm max}$ of the susceptibility maximum as a measure for the leading exchange $J$ in the spin chains, the following scenario was suggested:\cite{Cong2011} for Br substitution on the Cl1 and Cl2 sites, where $T_{\rm max} \approx 3$~K for $x<2$, the intra-chain exchange remains practically unchanged but increases significantly for substitution on the Cl3 sites, where for $x \geq 2$ the characteristic temperature increases linearly up to $T_{\rm max} \approx 9$~K at $x=4$. Very recently, the phase diagram has been extended down to lower temperatures by means of neutron scattering experiments.\cite{Well2018} Long range order was found to be suppressed for the intermediate concentration range $1.5 \leq x \leq 3.2$. In the same paper, accompanying density functional theory (DFT) calculations suggest a more complex evolution of the exchange integrals $J$ and $J^{\prime}$ with the bromine content $x$.

In this report we present an electron spin respnance (ESR) study of the substitutional series Cs$_2$CuCl$_{4-x}$Br$_x$. Previously we have shown that in Cs$_2$CuCl$_4$ the spin relaxation is governed by the uniform Dzyaloshinskii-Moriya (DM) interaction within the Cu--Cl chains due to the intra-chain Cu--Cl3--Cu bond angle which significantly deviates from $180^{\circ}$.\cite{Fayzullin2013} From the anisotropy of the resonance linewidth we determined the DM vector in good agreement with antiferromagnetic resonance investigations.\cite{Povarov2011} Already earlier the DM interaction was found to be essential for the understanding of the excitation spectra not only of Cs$_2$CuCl$_4$ but also of Cs$_2$CuBr$_4$.\cite{Fjaerestad2007} A few years ago, the exchange parameters of both compounds were determined by means of high-field ESR in the ordered phase.\cite{Zvyagin2014} Recently, paramagnetic resonance at high frequency in Cs$_2$CuBr$_2$ was analyzed \cite{Schulze2017} to determine the DM interaction following our approach in Cs$_2$CuCl$_2$.\cite{Schulze2017} Here we perform a comprehensive analysis of the anisotropy of the ESR linewidth in Cs$_2$CuCl$_{4-x}$Br$_x$ to determine the evolution of the DM vector dependent on the bromine substitution.

\section{Experimental Details}

High-quality single crystals of Cs$_2$CuCl$_{4-x}$Br$_x$ were grown from aqueous solution by an evaporation technique as described in Ref.~\onlinecite{Krueger2010} and characterized by magnetization and ultrasound measurements. \cite{Cong2011, Kreisel2011} ESR measurements were performed in a Bruker ELEXSYS E500 CW-spectrometer at X-band ($\nu \approx  9.36$~GHz) and Q-band frequencies ($\nu \approx  34$~GHz) equipped with continuous He-gas-flow cryostats (Oxford Instruments) in the temperature region $4 \leq T \leq 300$~K. The single crystals were fixed with well defined orientation in high-purity Suprasil quartz-glass tubes by paraffin and mounted at a computer-controlled goniometer allowing for angular dependent measurements. ESR spectra display the power $P$ absorbed by the sample from the transverse magnetic microwave field as a function of the static magnetic field $H$. The signal-to-noise ratio of the spectra is improved by recording
the derivative $dP/dH$ using lock-in technique with field modulation.

\section{Results}

\begin{figure}
\includegraphics[width = 0.8\columnwidth]{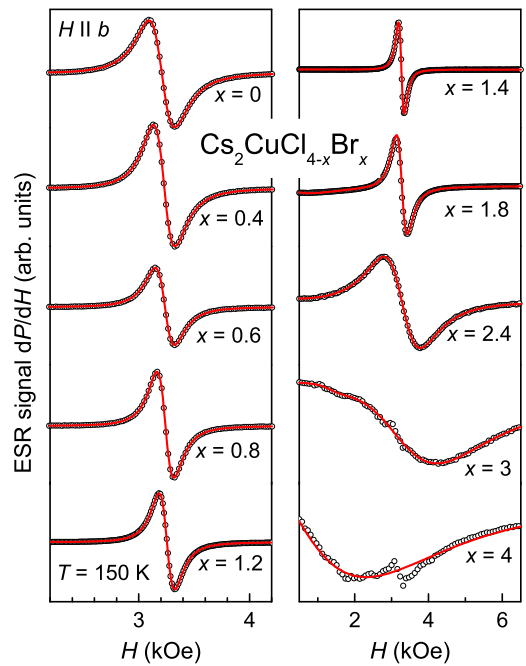}
\caption{(Color online) ESR spectra of Cs$_2$CuCl$_{4-x}$Br$_x$ at selected concentrations for the magnetic field $H$ applied along the crystallographic $b$ axis. The red solid lines indicate the fit by a Lorentzian line shape.} \label{spektrax0}
\end{figure}

In the whole temperature range and for all orientations of the magnetic field $H$ the ESR signal of Cs$_2$CuCl$_{4-x}$Br$_x$ consists of a single exchange-narrowed resonance line as exemplarily shown in
Fig.~\ref{spektrax0} for $H \parallel b$. The line is well fitted by a Lorentz shape at resonance field $H_{\rm res}$ with linewidth HWHM (half width at half maximum) $\Delta H$, indicating that spin-diffusion effects are not relevant in Cs$_2$CuCl$_4$.\cite{Dietz1971,Hennessey1973} For $x>1.2$ the linewidth significantly increases with increasing Br concentration, exceeding 2\,kOe for $x>3$  where the signal finally becomes too broad to be reasonably evaluated. For spectra, where the linewidth is of the same order of magnitude as the resonance field it was necessary to include the counter resonance at negative resonance field into the fitting to obtain a reasonable description of the resonance signal.\cite{Joshi2004} The $g$ value is obtained from resonance field and microwave frequency $\nu$ via the Larmor condition $h\nu=g\mu_{\rm B}H_{\rm res}$.

\begin{figure}
\includegraphics[width = 0.85\columnwidth]{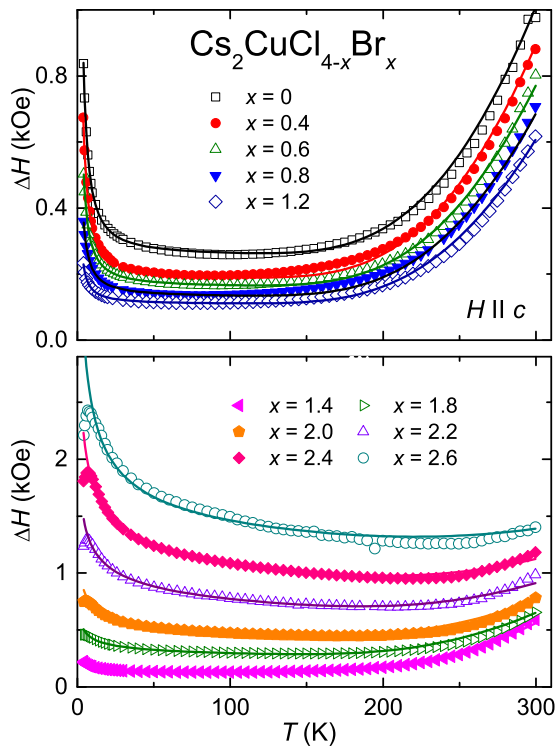}
\caption{(Color online) Temperature dependence of the linewidth $\Delta H$ of Cs$_2$CuCl$_{4-x}$Br$_x$ with different Br concentration $x$ for the magnetic field $H$ applied along the crystallographic $c$ axis. The solid lines indicate fits by Eq.~\ref{DH_T} containing an Arrhenius law and a critical divergence.}
\label{tempx1}
\end{figure}

\begin{figure}
\includegraphics[width=0.85\columnwidth]{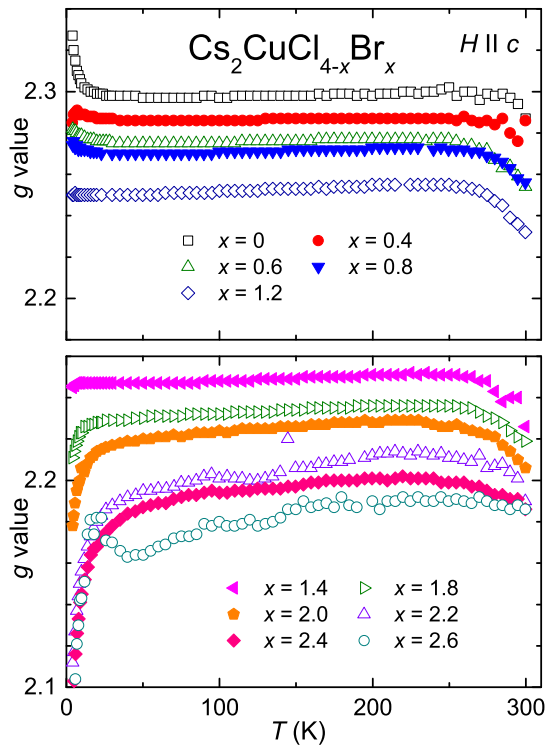}
\caption{(Color online) Temperature dependence of the $g$ value of Cs$_2$CuCl$_{4-x}$Br$_x$ with different $x$ for the magnetic field $H$ applied along the crystallographic $c$ axis.}
\label{tempx2}
\end{figure}

Before discussing the anisotropy of resonance linewidth $\Delta H$ and $g$ values in detail, we first focus on the temperature dependence, which is qualitatively similar for all orientations. The evolution of the temperature dependent linewidth and $g$ value with Br concentration $x$ is exemplarily shown in Figs.~\ref{tempx1} and \ref{tempx2}, respectively, for the magnetic field applied along the crystallographic $c$ axis.
Like in pure Cs$_2$CuCl$_4$ the linewidth exhibits three temperature regimes with different behavior. For all Br concentrations $x$ the linewidth is only weakly temperature dependent between 50\,K and 150\,K. It increases following an Arrhenius law $\Delta H \propto \exp{(-\Delta/T)}$ for high temperatures $T>150$~K. To low temperatures, for $T<50$~K the linewidth increases on approaching the ground state approximately by a power law $\Delta H \propto T^{-p}$.

The $g$ value is nearly constant and slightly larger than 2 as typical for Cu$^{2+}$ ions in the solid state, where the orbital moment is quenched by the crystal field. Only below 50\,K and above 250\,K temperature-dependent shifts of several percent occur, which are always negative for high temperatures -- here the anisotropy of the $g$ tensor diminishes due to thermal excitations -- but positive for $x < 0.8$, and again negative for $x > 0.8$ at low temperatures, indicating a change of local field distributions on formation of the ground states for increasing Br concentration.

\begin{figure}
\includegraphics[width=0.7\columnwidth]{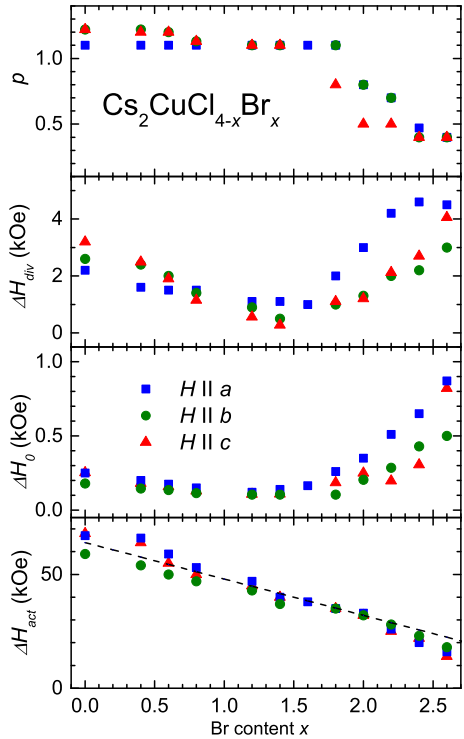}
\caption{(Color online) Concentration dependence of (from top to bottom) low-temperature critical exponent $p$ and corresponding divergent linewidth contribution $\Delta H_{\rm div}$ as well as constant contribution $\Delta H_0$ and activated contribution $\Delta H_{\rm act}$ with a gap value $\Delta = 1350$\,K. The parameters are obtained from fitting the linewidth of Cs$_2$CuCl$_{4-x}$Br$_x$ by Eq.~\ref{DH_T} for the magnetic field $H$ applied along the crystallographic $a$ (blue squares), $b$ (green circles) and $c$ axes (red triangles). The dashed line in the lower frame indicates a linear decrease of $\Delta H_{\rm act}$ on increasing $x$.}
\label{tempx3}
\end{figure}

For a quantitative evaluation we approximated the temperature dependence of the linewidth data for all three crystallographic axes by the following equation:
\begin{equation}
\Delta H = \Delta H_0 + \Delta H_{\rm act} \cdot \exp{(-\Delta/T)} + \Delta H_{\rm div}\left(\frac{1{\rm \,K}}{T}\right)^p \label{DH_T}
\end{equation}
Here $\Delta H_{0}$ denotes the high-temperature asymptotic contribution of pure spin-spin relaxation. The second term describes a thermally activated contribution $\Delta H_{\rm act}$ characterized by an energy gap $\Delta$, while the last term accounts for a divergent contribution $\Delta H_{\rm div}$ at low temperature with critical exponent $p$ normalized to $T=1$~K. As the scattering of the gap value $\Delta$ turned out to be larger than any systematic evolution with the Br content $x$, we chose a reasonable average value of $\Delta$~=~1350\,K as fixed parameter and varied only the four remaining parameters $\Delta H_{0}$, $\Delta H_{\rm act}$, $\Delta H_{\rm div}$ and $p$ in the fitting procedure. The corresponding results are shown in Figure~\ref{tempx3}. While the behavior of $\Delta H_{0}$, $\Delta H_{\rm div}$ and $p$ can be divided into two regimes $x < 1.6$ and $x > 1.6$, the parameter $\Delta H_{\rm act}$ decreases approximately linearly on increasing $x$, as indicated by the dashed line in the bottom frame of Fig.~\ref{tempx3}, approaching $\Delta H_{\rm act}(x=4) \rightarrow 0$. This indicates that the activated process is connected to the dynamics of the Cl$^{-}$ ions. In this respect it is important to note that the crystals show irreversible changes when heating above 370\,K. Thus, the energy gap probably is related to the structural stability of the compound. The critical exponent $p$ remains practically unchanged at $p \approx 1.1$ for $x \leq 1.6$, but decreases of $x \geq 1.6$ only. The simultaneous change of $\Delta$H$_{0}$ and $\Delta H_{\rm div}$ on increasing $x$, i.e. the monotonous decrease for $x \leq 1.6$  and the subsequent stronger increase for $x \geq 1.6$ suggest that both contributions to the linewidth arise from the Dzyaloshinskii-Moriya interaction.

\begin{figure}
\includegraphics[width=0.8\columnwidth]{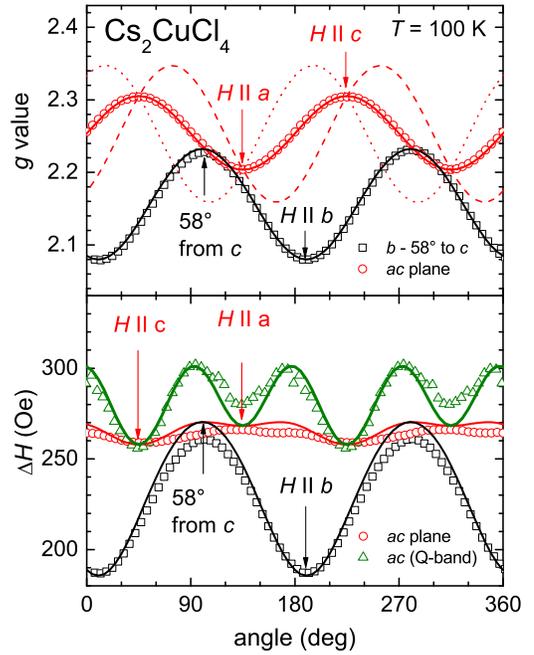}
\caption{(Color online) Angular dependence of $g$ value (upper frame) and linewidth $\Delta H$ (lower frame) in Cs$_2$CuCl$_{4}$ at $T = 100$\,K for the magnetic field applied within the $ac$ plane and a perpendicular plane containing the $b$ axis. Black squares and red circles: X-band, green triangles: Q-band. Dashed and dotted lines indicate the two inequivalent $g$ tensors, solid lines are fits as described in the text.}
\label{fit0_0}
\end{figure}

\begin{figure}
\includegraphics[width=0.8\columnwidth]{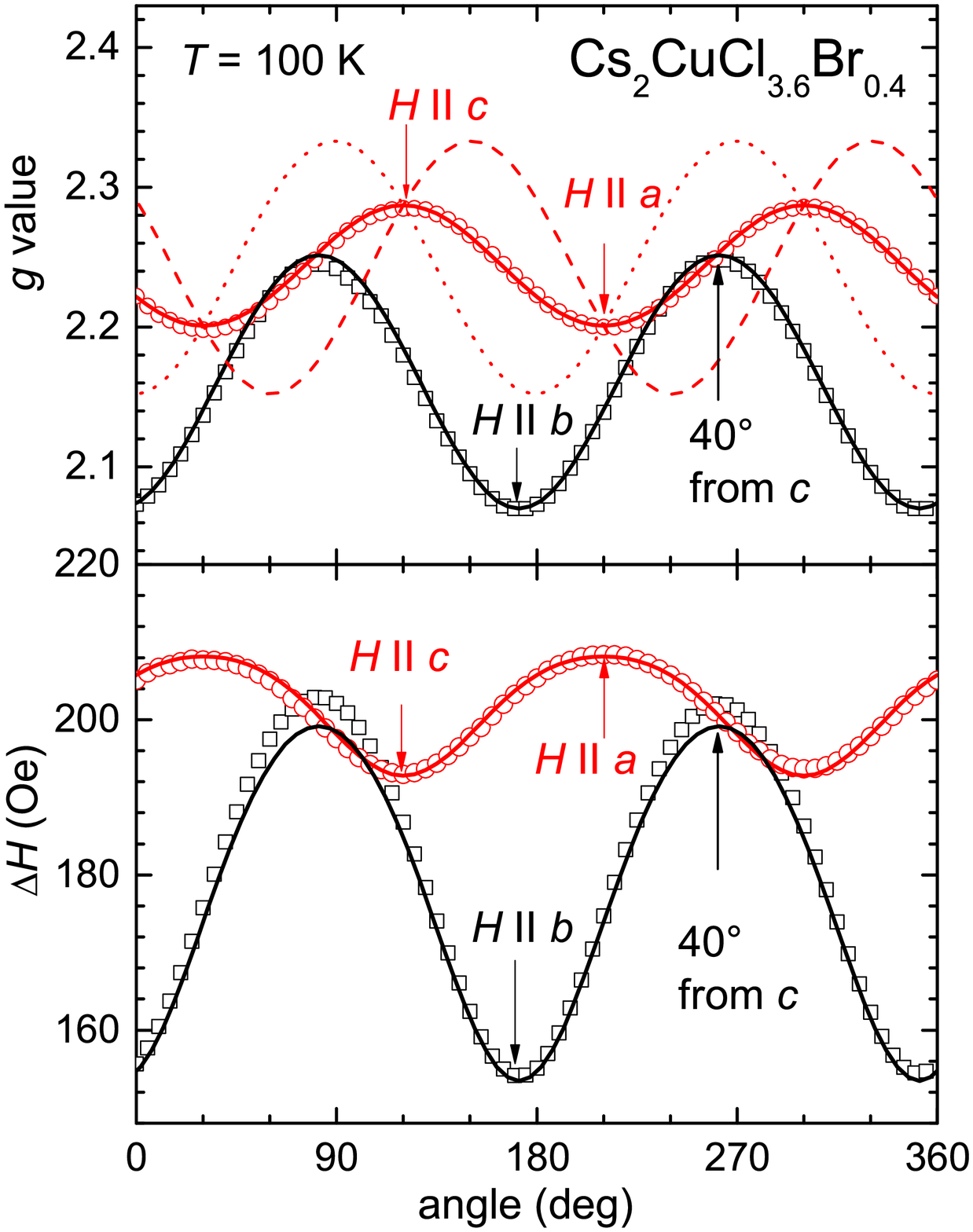}
\caption{(Color online) Angular dependence of g-value (upper frame) and linewidth $\Delta H$ (lower frame) in Cs$_2$CuCl$_{3.6}$Br$_{0.4}$ at $T = 100$\,K and X-band frequency for the magnetic field applied within the $ac$ plane and a perpendicular plane containing the $b$ axis. Dashed and dotted lines indicate the two inequivalent $g$ tensors, solid lines are fits as described in the text.}
\label{fit0_4}
\end{figure}

\begin{figure}
\includegraphics[width=0.8\columnwidth]{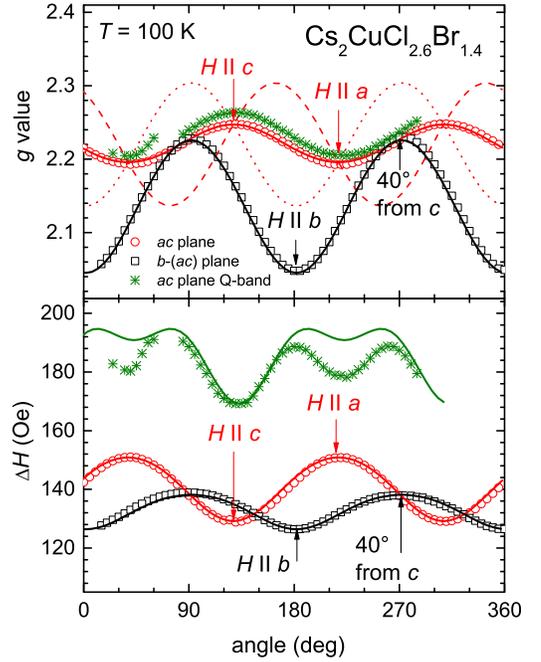}
\caption{(Color online) Angular dependence of g-value (upper frame) and linewidth $\Delta H$ (lower frame) in Cs$_2$CuCl$_{2.6}$Br$_{1.4}$ at $T = 100$\,K for the magnetic field applied within the $ac$ plane and a perpendicular plane containing the $b$ axis. Black squares and red circles: X-band, green stars: Q-band. The Q-band data have been measured on a different crystal with slightly larger linewidth. Dashed and dotted lines indicate the two inequivalent $g$ tensors, solid lines are fits as described in the text.}
\label{fit1_4}
\end{figure}

\begin{figure}
\includegraphics[width=0.8\columnwidth]{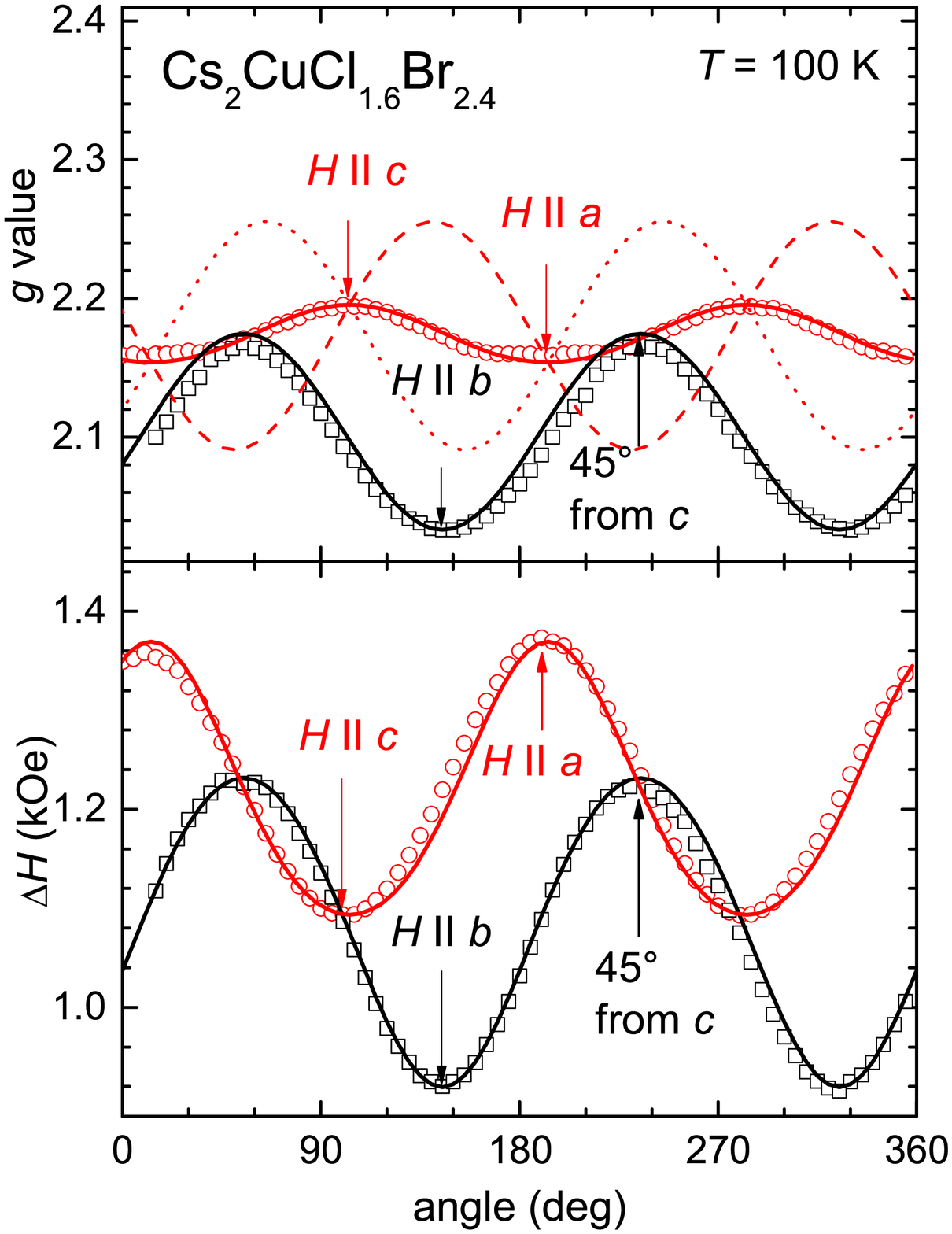}
\caption{(Color online) Angular dependence of g-value (upper frame) and linewidth $\Delta H$ (lower frame) in Cs$_2$CuCl$_{1.6}$Br$_{2.4}$ at $T = 100$\,K and X-band frequency for the magnetic field applied within the $ac$ plane and a perpendicular plane containing the $b$ axis. Dashed and dotted lines indicate the two inequivalent $g$ tensors, solid lines are fits as described in the text.}
\label{fit2_4}
\end{figure}

In the following we will show that the Dzyaloshinskii-Moriya interaction systematically changes with the substitution of Cl by Br, where first the sites outside of the Cu--Cl--Cu chain are substituted and only for large $x$ the sites within the chain become involved. For this purpose we measured the angular dependence of the $g$-value and linewidth at $T=100$\,K for all Br concentration under consideration. At this intermediate temperature the line broadening can be treated in high~temperature approximation, i.e $k_{B}T \gg J$, but the activated process can still be neglected. Figures ~\ref{fit0_0}--\ref{fit1_4} show exemplarily the anisotropies of selected Br concentrations $x$ for the magnetic field applied within the $ac$ plane (red circles) and in a plane containing the $b$ axis (black squares). These geometries have been chosen, because crystals with well defined orientation could be best prepared as platelets cut perpendicular to the chain $b$ axis, or as thin rods grown along the $b$ axis. From the coincidence of both data sets of linewidth and $g$ value one obtains the angle of the rotation plane containing the $b$ axis with respect to the $c$ axis as indicated in Figs.~\ref{fit0_0}--\ref{fit2_4}.

\section{Analysis and Discussion}

To evaluate the anisotropy of the ESR data, the relevant linewidth contributions due to DM interaction and anisotropic Zeeman effect have been derived previously \cite{Fayzullin2013} and refined recently.\cite{Schulze2017} Here we only shortly describe the main steps of the calculation and final expressions.

To derive the contribution of the uniform DM interaction to the linewidth in Cs$_2$CuCl$_4$, one starts from the one dimensional Heisenberg Hamiltonian
\begin{equation}
{\cal H} = \sum_i J \mathbf{S}_i\cdot\mathbf{S}_{i+1} + \sum_i \mathbf{D}\cdot[\mathbf{S}_i \times \mathbf{S}_{i+1}] + {\cal H}_{\rm ic}(J^{\prime}) \label{Hamiltonian}
\end{equation}
describing the Cu$^{2+}$ spin $S=1/2$ chains along the crystallographic $b$ axis. Here $J$ is the isotropic antiferromagnetic intra-chain superexchange coupling parameter between nearest neighbor spins $\mathbf{S}_{i}$ and $\mathbf{S}_{i+1}$ within the chains and $\bf{D}$ denotes the intra-chain DM vector. The contribution ${\cal H}_{\rm ic}(J^{\prime})$ takes the isotropic inter-chain exchange $J^{\prime}$ into account, which couples each spim ${\bf S}_i$ with two spins on both adjacent chains within the $bc$ plane.

In the crystal structure, there exist four inequivalent chains which differ in the orientation of the corresponding DM vector $\bf{D}$. For their description we follow the notations of Starykh \textit{et al.}.\cite{Starykh2010} The laboratory coordinate system $xyz$ is chosen with the $z$-axis along the externally applied magnetic field. In order to apply the usual expressions for the transformation of the $\bf{D}$ components between laboratory $(xyz)$ and crystallographic $(XYZ)$ coordinate systems the following notations were used: The $Y$ axis is chosen along the chain (i.e. crystallographic $b$ axis), $Z$- and $X$-axis are parallel to $c$ and $a$, respectively. Then the components of the DM vector transform following
\begin{eqnarray}
D^x &=& D^X\cos{\beta}\cos{\alpha}+D^Y\cos{\beta}\sin{\alpha}-D^Z\sin{\beta}, \nonumber \\
D^y &=& D^Y\cos{\alpha}-D^X\sin{\alpha}, \label{Dtransform} \\
D^z &=& D^X\sin{\beta}\cos{\alpha}+D^Y\sin{\beta}\sin{\alpha}+D^Z\cos{\beta}, \nonumber
\end{eqnarray}
with
\begin{equation}
\cos\alpha=\frac{A}{\sqrt{A^2+B^2}}, \cos\beta=\frac{C}{\sqrt{A^2+B^2+C^2}},\nonumber
\end{equation}
and the $g$ value given by
\begin{equation}
g = \sqrt{A^2+B^2+C^2},\nonumber
\end{equation}
where
\begin{eqnarray}
A&=&g_{aa}\sin\theta\cos\phi+g_{ab}\sin\theta\sin\phi+g_{ac}\cos\theta, \nonumber\\
B&=&g_{ba}\sin\theta\cos\phi+g_{bb}\sin\theta\sin\phi+g_{bc}\cos\theta, \label{Dtransform1} \\
C&=&g_{ca}\sin\theta\cos\phi+g_{cb}\sin\theta\sin\phi+g_{cc}\cos\theta. \nonumber
\end{eqnarray}
Here, the angles $\alpha$ and $\beta$  define the orientation of the local coordinate system in which the Hamiltonian of the Zeeman energy
${\cal H}_i^{\rm Zee} = \mu_{B}\mathbf{H} \mathbf{g} \mathbf{S}_i$
takes diagonal form, while $\theta$ is the polar angle between external magnetic field and $c$ axis, and $\phi$ is the azimuthal angle counted from the $a$ direction. $g_{\xi,\eta}$ with $\xi,\eta = a,b,c$ denote the components of the $\mathbf{g}$ tensor in the crystallographic coordinate system.
These transformations should be applied for all four chains using the corresponding settings $D_1^X=D_a$, $D_1^Z=-D_c$ in chain 1, $D_2^X=-D_a$, $D_2^Z=-D_c$ in chain 2,
$D_3^X=-D_a$, $D_3^Z=D_c$ in chain 3, and $D_4^X=D_a$, $D_4^Z=D_c$ in chain 4, following the notation of Ref.~\onlinecite{Starykh2010}.

Using the theory of exchange narrowing, in the high-temperature limit $(k_{\rm B} T \gg J)$ the angular dependence of the half-width at half maximum $\Delta H$ of a single chain is determined by\cite{Kubo1954,Fayzullin2013}
\begin{equation}
\Delta H_{\rm DM} = C \frac{D^2(\theta,\phi)k_{\rm B}}{\sqrt{\langle J^2 \rangle} g\mu_{\rm B}} \frac{1}{2\sqrt{2}} \label{DHinf}
\end{equation}
The parameters $D$ and $J$ are taken in Kelvin, $\mu_{\rm B}$ and $k_{\rm B}$ denote Bohr magneton and Boltzmann constant, respectively. The prefactor $C$ depends on the cut-off conditions for the spectra and is taken as $C=\pi/\sqrt{2}$ assuming exponential wings. The brackets $\langle J^2 \rangle$ indicate the averaged squared isotropic exchange integral including inter-chain interactions as described below.

\begin{figure}
\includegraphics[width=0.8\columnwidth]{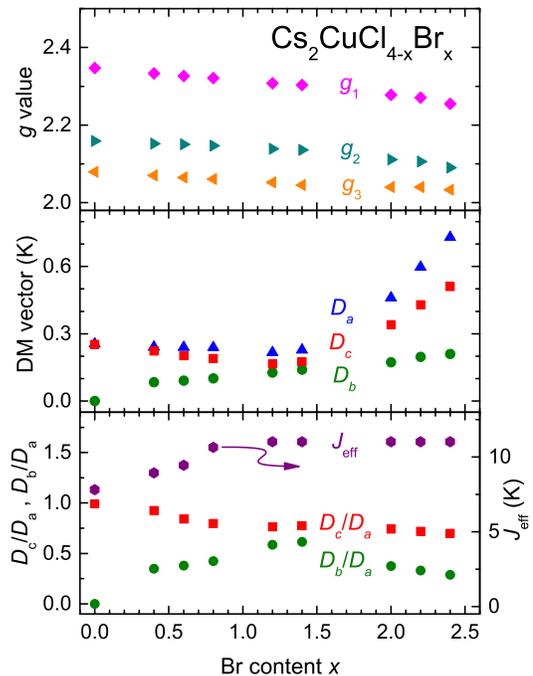}
\caption{(Color online) Concentration dependence of the diagonal components of the $g$ tensor $(upper frame)$ in local coordinates, DM-vector components $(middle frame)$ and ratio of DM components as well as effective exchange parameter $J_{\rm eff} = \sqrt{\langle J \rangle^2}$ $(lower frame)$ of Cs$_2$CuCl$_{4-x}$Br$_x$.}
\label{concent}
\end{figure}

Note that there are two magnetically nonequivalent Cu$^{2+}$ positions with different orientation of the $\mathbf{g}$ tensor within the $ac$ plane, i.e. $\mathbf{g}_{I}$ and $\mathbf{g}_{II}$, in the unit cell, but within a single chain all $\mathbf{g}$ tensors are equivalent. Details can be found in Ref.~\onlinecite{Fayzullin2013}. The effective $g$ value for a certain magnetic field direction is given by the arithmetic average $g=(g_{I}+g_{II})/2$ of the local $g$ values of the two inequivalent Cu sites in the crystal structure (dashed and dotted lines in the upper frames of Figs.~\ref{fit0_0}--\ref{fit1_4}). Supporting measurements of the linewidth anisotropy in the $ac$ plane have been performed at Q--band frequency for some of the crystals to cross--check the local $g$ tensors based on the line broadening due to the anisotropic Zeeman effect, given by
\begin{equation}
\Delta H_{\rm AZ} = \left(\frac{g_I-g_{II}}{g}\right)^2 \frac{g \mu_{\rm B} H_{\rm res}^2}{\sqrt{\langle J^2 \rangle}} \label{DHAZ}
\end{equation}
which results in the 90$^{\circ}$ modulation (green triangles in the lower frames shown in Figures~\ref{fit0_0} and ~\ref{fit1_4}). Here $\langle J^2 \rangle = (J^2+2(J')^2)/3$ denotes the averaged squared exchange integral consisting of one nearest neighbor contribution $J$ within the chain and two links $J^{\prime}$ to the neighboring chain due to the triangular structure. The dependence of the local $\mathbf{g}$ tensor on the Br content is depicted in the top frame of Fig.~\ref{concent}. After refined evaluation of the data the absolute values of the local $g$ tensor components obtained for $x=0$ have been improved as compared to those values obtained earlier in Ref.~\onlinecite{Fayzullin2013}. The principal values simultaneously slightly decrease on increasing $x$, while the sequence g$_{1}>$~g$_{2}>$~g$_{3}$ remains conserved. 

To describe the experimentally observed angular dependence of the linewidth, we summed up the DM contributions Eq.~\ref{DHinf} of all four inequivalent chains plus the anisotropic Zeeman contribution Eq.~\ref{DHAZ} due to the different orientation of the $\textbf{g}$ tensors of the two inequivalent copper sites. For the perpendicular plane the anisotropic Zeeman contribution is of minor importance. From the angular dependence of the linewidth we obtained the components of the DM vector as well as the effective exchange parameter $J_{\rm eff} = \sqrt{\langle J \rangle^2}$, respectively. Note that regarding Eqs.~\ref{DHinf} and \ref{DHAZ}, only the sum of the squared exchange parameters accounts for the exchange narrowing of the ESR signal. Therefore, we obtain only absolute values of the effective exchange parameter. Thus, we are not sensitive to a change of sign, i.e. we cannot distinguish between ferro- and antiferromagnetic exchange. Moreover, without further assumptions it is impossible to distinguish a change in $J^{\prime}$ from a change in $J$. Here we use the information obtained from the $x$ dependence of $T_{\rm max}$ in the magnetic susceptibility.\cite{Cong2011}

The evolution of the DM vector with increasing $x$ is shown in the middle frame of Figure ~\ref{concent}. For $x=0$, $D_{a} \simeq D_{c}$ while the component $D_{b}=0$ in agreement with the results obtained in Ref.~\onlinecite{Fayzullin2013}. For $x < 1.6$ $D_{a}$ and $D_{c}$ change only slightly, but $D_{b}$ becomes non zero and gradually increases on increasing $x$. For $x > 1.6$ $D_{a}$ and $D_{c}$ rapidly increase, while $D_{b}$ continues with the same slope as for $x < 1.6$. The observed changes are in line with the site selective substitution of Cl by Br, $D_{a}$ and $D_{c}$ depend on the Cu--Cl--Cu bond within the chains. Therefore, they remain nearly unchanged at low Br content $x$ when Br substitutes only Cl sites out of the chains, but strongly increase for substitution within the chains. Notably the effective exchange $J_{\rm eff}$ increases for $0 \leq x \leq 1$ from about 7~K to 11~K, due to the increase of the inter-chain coupling $J^{\prime}$ on substitution of Cl1 sites by Br from $J^{\prime} = 0.3 J$ for $x=0$ to $J^{\prime} = 0.8 J$ for $x=1$. In the same range of $x$ the ratio $D_c/D_a$ decreases from 1 down to 0.75. For further increasing $x$ both quantities change only slightly when the Cl2 sites are substituted by Br. For $x > 2.4$ where the linewidth became too large to be analyzed, an increase of $J_{\rm eff}$ is expected, as the Br ions substituted within the chains obviously strongly increase the intra-chain exchange. Such an increase of the intra-chain exchange is in agreement with the increase of the temperature, where the characteristic maximum of the spin-chain susceptibility is located, from 3 K at $x=1.6$ up to 9 K for $x=4$ as documented in Ref.~\onlinecite{Cong2011}. This is also in accordance with comparative antiferromagnetic resonance experiments \cite{Zvyagin2014} on the pure compounds $x=0$ amd $x=4$ yielding $J(x=4)/J(x=0) \approx 3$ and $J^{\prime}(x=4)/J^{\prime}(x=0) \approx 4$. Regarding the DM interaction, for $x > 2.4$ the ratio $D_c/D_a$ is expected to decrease gradually down to the value $D_c/D_a=0.3$ reported for the pure Br compound.\cite{Zvyagin2014,Schulze2017} Finally the ratio $D_b/D_a$, which increases for $x<1.6$ but decreases for larger $x$, suggests that the component $D_b$ of the DM vector results from random local distortions of the bond geometry when gradually replacing Cl by Br and, therefore, vanishes for the pure compounds at $x=0$ and $x=4$.

Note that our present results are also in qualitative agreement with recent DFT calculations performed for $x=0$, 1, 2, and 4, where the exchange ratio $J^{\prime}/J$ is also significantly increasing for $0 \leq x \leq 1$ due to an increase of the inter-chain exchange.\cite{Well2018} In those calculations, which assumed 100\% site selectivity of the Br substitution, the inter-chain exchange $J^{\prime}$ becomes even larger than the intra-chain exchange $J$, which itself switches its sign to ferromagnetic interaction for $x=1$ and $x=2$. Moreover an additional ferromagnetic inter-layer exchange $J^{\prime\prime}$ contribution turns out to become sizable. A direct application of the DFT results to evaluate the ESR data imposes the question of correct interpolation between the calcutated concentrations $x = 0$, 1, 3, and 4, which consider only ideally site-ordered systems. A certain degree of disorder can be expected to influence the results. Especially ferromagnetic super exchange is known to be extremely sensitive to slight distortions. In any case to estimate the exchange parameters for intermediate bromine contents, it is necessary to analyze the effect of random occupations of the Cl sites by Br. Nevertheless regarding  the existing DFT results, the square root of the sum of the squared exchange contributions remains of the order of 1~meV, providing a stable strength of exchange-narrowing of the ESR signal. Using the exchange-values of the DFT calculations may slightly alter the absolute values of the DM vector, but in any case the orientation of the DM vector remains unaffected.

\section{Conclusion}

Our comprehensive electron-spin-resonance investigation of the substitutional series Cs$_2$CuCl$_{4-x}$Br$_x$ provides important microscopic information on the evolution of local exchange couplings and relaxation channels in this frustrated quantum-spin $S=1/2$ chain compound. The simultaneous evaluation of the anisotropy of $g$ value and linewidth taking into account the anisotropic Zeeman effect allows to determine the anisotropic exchange contributions. Starting from $x=0$
the Dzyaloshinskii-Moriya (DM) interaction provides the dominant relaxation channel with the DM vector oriented within the $ac$ plane with an angle of about $45^{\circ}$ with respect to the $a$ axis. On increasing $x$, the DM vector gradually rotates to approximately $35^{\circ}$ within the $ac$ plane for high $x$. The absolute value of the DM vector changes only slightly for $x<1.6$ but increases significantly for $x>1.6$, when Br ions mainly substitutes Cl ions within the chains.

\acknowledgments

We are grateful to N.~van~Well (Universit\"at Bayreuth) for crystal growth. We thank C.~Krellner and M.~Lang (Goethe Universit\"at Frankfurt) and J. Deisenhofer (Augsburg University) for useful discussions. We acknowledge financial support by the Deutsche Forschungsgemeinschaft (DFG) via the Transregional Research Center TRR\,80 (Augsburg, M\"unchen, Stuttgart) and SFB/TR\,49 (Frankfurt). R.M.E.'s work was done within the framework of fundamental research of FRC Kazan Scientific Center of RAS.


\begin{thebibliography}{20}

\bibitem{Vasiliev2018} A. Vasiliev, O. Volkova, E. Zvereva and M. Markina, \textit{Quantum Materials} \textbf{3}, 18 (2018).

\bibitem{Balents2010} L. Balents, \textit{Nature} \textbf{464}, 199 (2010).

\bibitem{Mellor1939} D. P. Mellor, \textit{Z. Krist.} \textbf{101}, 160 (1939).

\bibitem{Helmholz1952} L. Helmholz and R. F. Kruh, \textit{J. Am. Chem. Soc.} \textbf{74}, 1176 (1952).

\bibitem{Morosin1960} B. Morosin and E. C. Lingafelter, \textit{Acta Crystallogr.} \textbf{13}, 807 (1960).

\bibitem{Coldea2001} R. Coldea, D. A. Tennant, A. M. Tsvelik, and Z. Tylczynski, \textit{Phys. Rev. Lett.} \textbf{86}, 1335 (2001).

\bibitem{Starykh2007} Oleg A. Starykh and Leon Balents, \textit{Phys. Rev. Lett.} \textbf{98}, 077205 (2007).


\bibitem{Ono2004} T. Ono, H. Tanaka, O. Kolomiyets, H. Mitamura, T. Goto, K. Nakajima, A. Oosawa, Y. Koike, K. Kakurai, J. Klenke, P Smeibidle and M Meißner, \textit{J. Phys.: Condens. Matter} \textbf{16}, S773 (2004).


\bibitem{Ono2003} T. Ono, H. Tanaka, H. Aruga Katori, F. Ishikawa, H. Mitamura, and T. Goto, \textit{Phys. Rev. B} \textbf{67}, 104431 (2003).

\bibitem{Coldea2002}  R. Coldea, D. A. Tennant, K. Habicht, P. Smeibidl, C. Wolters, and Z. Tylczynski, \textit{Phys. Rev. Lett.} \textbf{88}, 137203 (2002).

\bibitem{Radu2005} T. Radu, H. Wilhelm, V. Yushankhai, D. Kovrizhin, R. Coldea, Z. Tylczynski, T. Luhmann, and F. Steglich, \textit{Phys. Rev. Lett.} \textbf{95}, 127202 (2005).

\bibitem{Vachon2011} M.-A. Vachon, G. Koutroulakis, V. F. Mitrovic, O. Ma, J. B. Marston, A. P. Reyes, P. Kuhns, R. Coldea, and Z. Tylczynski,  \textit{New J. Phys.} \textbf{13}, 093029 (2011).

\bibitem{Smirnov2012} A. I. Smirnov, K. Y. Povarov, S. V. Petrov, and A. Y. Shapiro, \textit{Phys. Rev. B} \textbf{85}, 184423 (2012).

\bibitem{Smirnov2015} A. I. Smirnov, T. A. Soldatov, K. Yu. Povarov, and A. Ya. Shapiro, \textit{Phys. Rev. B} \textbf{91}, 174412 (2015).

\bibitem{Streib2015} S. Streib, P. Kopietz, P. T. Cong, B. Wolf, M. Lang, N. van Well, F. Ritter, and W. Assmus, \textit{Phys. Rev B} \textbf{91}, 041108(R) (2015).

\bibitem{Cong2016} P. T. Cong, L. Postulka, B. Wolf, N. van Well, F. Ritter, W. Assmus, C. Krellner, and M. Lang, \textit{J. Appl. Phys.} \textbf{120}, 142113 (2016).

\bibitem{Kim2017} Hyeong-Jin Kim, C. R. S. Haines, C. Liu, Sae Hwan Chun, Kee Hoon Kim, H. T. Yi, Sang-Wook Cheong, and Siddharth S. Saxena, \textit{Low Temp. Phys.} \textbf{43}, 901 (2017).

\bibitem{Fujii2007} Y. Fujii, H. Hashimoto, Y. Yasuda, H. Kikuchi, M. Chiba, S. Matsubara and M. Takigawa, \textit{J. Phys.: Condens. Matter} \textbf{19}, 145237 (2007).

\bibitem{Alicea2009} J. Alicea, A. V. Chubukov, and O. A. Starykh, \textit{Phys. Rev. Lett.} \textbf{102}, 137201 (2009).

\bibitem{Wolf2013} B. Wolf, P. T. Cong, N. Kr\"uger, F. Ritter, W. Assmus, and M. Lang, \textit{J. Low Temp. Phys.} \textbf{170}, 236 (2013).

\bibitem{Zvyagin2015} S. A. Zvyagin, M. Ozerov, D. Kamenskyi, J. Wosnitza, J. Krzystek, D. Yoshizawa, M. Hagiwara, R. Hu, H. Ryu, C. Petrovic, and M. E. Zhitomirsky, \textit{New J. Phys.} \textbf{17}, 113059 (2015).

\bibitem{Cong2011} P. T. Cong, B. Wolf, M. de Souza, N. Kr\"uger, A. A. Haghighirad, S. Gottlieb-Schoenmeyer, F. Ritter, W. Assmus, I. Opahle, K. Foyevtsova, H. O. Jeschke, R. Valenti, L. Wiehl, and M. Lang, \textit{Phys. Rev. B} \textbf{83}, 064425 (2011).

\bibitem{Well2018} N. van Well, O. Zaharko, B. Delley, M. Skoulatos, R. Georgii, S. van Smaalen, and Ch. R\"uegg, \textit{Ahh. Phys.} \textbf{2018}, 1800270 (2018).

\bibitem{Fayzullin2013} M. A. Fayzullin, R. M. Eremina, M. V. Eremin, A. Dittl, N. van Well, F. Ritter, W. Assmus, J. Deisenhofer, H.-A. Krug von Nidda, and A. Loidl, \textit{Phys. Rev. B} \textbf{88}, 174421 (2013).

\bibitem{Povarov2011} K. Y. Povarov, A. I. Smirnov, O. A. Starykh, S. V. Petrov, and A. Y. Shapiro, \textit{Phys. Rev. Lett.} \textbf{107}, 037204 (2011).

\bibitem{Fjaerestad2007}J. O. Fjaerestad, W. Zheng, R. R. P. Singh, R. H. McKenzie, and R. Coldea, \textit{Phys. Rev. B} \textbf{75}, 174447 (2007).

\bibitem{Schulze2017} E. Schulze, A. N. Ponomaryov, J. Wosnitza, H. Tanaka, and S. A. Zvyagin \textit{Low Temp. Phys.} \textbf{43}, 1311 (2017).

\bibitem{Zvyagin2014} S. A. Zvyagin, D. Kamenskyi, M. Ozerov, J. Wosnitza, M. Ikeda, T. Fujita, M. Hagiwara, A. I. Smirnov, T. A. Soldatov, A. Ya. Shapiro, J. Krzystek, R. Hu, H. Ryu, C. Petrovic, and M. E. Zhitomirsky, \textit{Phys. Rev. Lett.} \textbf{112}, 077206 (2014).

\bibitem{Krueger2010} N. Krueger, S. Belz, F. Schossau, A. A. Haghighirad, P. T. Cong, B. Wolf, S. Gottlieb-Schoenmeyer, F. Ritter, and W. Assmus, \textit{Cryst. Growth Des.} \textbf{10}, 4456 (2010).

\bibitem{Kreisel2011} A. Kreisel, P. Kopietz, P. T. Cong, B. Wolf, and M. Lang, \textit{Phys. Rev. B} \textbf{84}, 024414 (2011).

\bibitem{Dietz1971} R. E. Dietz, F. R. Merrit, R. Dingle, D. Hone, B. G. Silbernagel, and P. M. Richards, \textit{Phys. Rev. Lett.} \textbf{26}, 1186 (1971).

\bibitem{Hennessey1973} M. J. Hennessey, C. D. McElwee, and P. M. Richards, \textit{Phys. Rev. B} \textbf{7}, 930 (1973).

\bibitem{Joshi2004} J. P. Joshi and S. V. Bhat, \textit{J. Magn. Reson.} \textbf{168}, 284 (2004).

\bibitem{Kubo1954} R. Kubo and K. Tomita, \textit{J. Phys. Soc. Jpn.} \textbf{9}, 888 (1954).

\bibitem{Castner1971} T. G. Castner and M. Seehra, \textit{Phys. Rev. B} \textbf{4}, 38 (1971).

\bibitem{Huber1999} D. L. Huber, G. Alejandro, A. Caneiro, M. T. Causa, F. Prado, M. Tovar, and S. B. Oseroff, \textit{Phys. Rev. B}  \textbf{60}, 12155 (1999).

\bibitem{Starykh2010} O. A. Starykh, H. Katsura, and L. Balents, \textit{Phys. Rev. B} \textbf{82}, 014421 (2010).

\end{thebibliography}
\end{document}